\documentclass[pra, amsmath, amssymb, superscriptaddress, reprint]{revtex4-1}%

\usepackage{dcolumn}
\usepackage{bm}
\usepackage{amsmath}
\usepackage{amsfonts}
\usepackage{amssymb}%
\usepackage{latexsym}
\usepackage{braket}
\usepackage{comment}
\usepackage{grffile}

\usepackage{graphicx}

\usepackage{color}

\begin{document}

\newcommand{\change}[2]{{\color{blue} #1}{\color{red} #2}}

\preprint{{\large Preprint}} 

\title{Application of the time-dependent surface flux method to the time-dependent multiconfiguration self-consistent-field method}

\author{Yuki Orimo}
\email{ykormhk@atto.t.u-tokyo.ac.jp}
  	\affiliation{Department of Nuclear Engineering and Management, Graduate School of Engineering, The University of Tokyo, 7-3-1 Hongo, Bunkyo-ku, Tokyo 113-8656, Japan}

\author{Takeshi Sato}
\author{Kenichi L. Ishikawa}
  	\affiliation{Department of Nuclear Engineering and Management, Graduate School of Engineering, The University of Tokyo, 7-3-1 Hongo, Bunkyo-ku, Tokyo 113-8656, Japan}
    \affiliation{Photon Science Center, Graduate School of Engineering, The University of Tokyo, 7-3-1 Hongo, Bunkyo-ku, Tokyo 113-8656, Japan}
  	\affiliation{Research Institute for Photon Science and Laser Technology, The University of Tokyo, 7-3-1 Hongo, Bunkyo-ku, Tokyo, 113-0033 Japan}



\begin{abstract}
We present a numerical implementation of the time-dependent surface flux (tSURFF) method [New J. Phys. 14, 013021 (2012)],
an efficient computational scheme to extract photoelectron energy spectra, 
to the time-dependent multiconfiguration self-consistent-field (TD-MCSCF) method. 
Extending the original tSURFF method developed for single
particle systems, we formulate the equations of motion for the spectral amplitude of orbital functions 
constutiting the TD-MCSCF wave function,
from which the angle-resolved photoelectron energy spectrum, and more generally, photoelectron reduced density matrices (RDMs) are readiliy 
obtained. The tSURFF method applied to the TD-MCSCF wave function, in combination with an efficient
absorbing boundary offered by the infinite-range exterior complex scaling, enables accurate {\it ab initio} computations of photoelectron energy spectra 
from multielectron systems subject to an intense and ultrashort laser pulse with a computational cost significantly reduced compared to that required in
projecting the total wave function onto scattering states. 
We apply the present implementation to the photoionization of Ne exposed to an attosecond extreme-ultraviolet (XUV) pulse and above-threshold ionization of Ar irradiated by an intense mid-infrared laser field, demonstrating both accuracy and efficiency of the present method.
\end{abstract}

\maketitle

\section{introduction}
\label{sub:introduction}

The rapid progress in experimental techniques for high-intensity, ultrashort optical pulses, has lead to the advent and development of strong-field physics and attosecond science \cite{Agostini_2004,Krausz_2009}, with the ultimate goal to directly measure and control electron motion in atoms, molecules, and solids.
Although the time-dependent Schr\"odinger equation (TDSE) provides a rigorous theoretical framework to investigate electron dynamics, solving it for multielectron systems poses a major challenge.
To simulate multielectron dynamics in intense laser fields, the time-dependent multiconfiguration self-consistent field (TD-MCSCF) methods have been developed \cite{Ishikawa_2015,Zanghellini_2003,Kato_2004,Caillat_2005,Nguyen-Dang_2007,Miyagi_2013,Sato_2013,Miyagi_2014,Haxton_2015,Sato_2015} which express the total wave function as a superposition,
\begin{equation}
    \label{eq:multiconfiguration expansion}
    \Psi (t) = \sum_I\Phi_I(t)C_I(t),
\end{equation}
of the Slater determinants (or electronic configurations) $\Phi_I(t)$ constructed from single-electron orbitals. Both the configuration-interaction (CI) coefficients $\{C_I\}$ and orbitals are varied in time.
If we consider all the possible configurations for a given number of orbitals as in the multiconfiguration time-dependent Hartree-Fock (MCTDHF) method \cite{Zanghellini_2003,Kato_2004,Caillat_2005}, the computational cost increases factorially with the number of electrons.
To overcome this difficulty, we have recently formulated and numerically implemented the time-dependent complete-active-space self-consistent-field (TD-CASSCF) method \cite{Sato_2013}, and even more general and less demanding time-dependent occupation-restricted multiple-active-space (TD-ORMAS) method \cite{Sato_2015}.
The former introduces orbital classification into doubly occupied core orbitals and fully correlated active orbitals; the core orbitals are further divided into time-independent frozen core and time-dependent dynamical core. 
The latter further divides active orbitals into an arbitrary number of subgroups, specifying the minimum and maximum numbers of electrons distributed in each subgroup. 
Flexible description of the wave function offered by the orbital classification and occupation restriction enables converged simulations of highly nonlinear, correlated multielectron dynamics in systems containing several tens of electrons.

Photoelectron energy spectra (PES) and their angular distribution, or angle-resolved photoelectron energy spectra (ARPES) are among important experimental probes.
In principle, they could be calculated by projecting the departing photoelectron wave packet onto plane waves or Coulomb waves.
This approach, however, requires retaining the complete wave function without being absorbed, leading to a huge simulation box and prohibitive computational cost.
To circumvent this difficulty, Tao and Scrinzi have devised the time-dependent surface flux (tSURFF) method \cite{Tao_2012}, which extracts PES by integrating the electron flux through a surface.
Thus it allows one to use an absorbing boundary, bringing significant cost reduction.
The tSURFF method was first developed for single electron systems \cite{Tao_2012} and then applied to multielectron simulations with, e.g., the hybrid antisymmetrized coupled channels method \cite{Majety_2015}, the time-dependent configuration interaction singles method \cite{Karamatskou_2014}, and the time-dependent density functional theory\cite{Wopperer_2017}.

In this study, we combine the tSURFF method with the TD-CASSCF and TD-ORMAS methods in order to extract photoelectron energy spectra from {\it ab initio} simulations of multielectron dynamics in atoms subject to intense and/or ultrashort laser pulses. Under a physically reasonable assumption that the nuclear potential and interelectronic Coulomb interaction are negligible for photoelectron dynamics in the region distant from the nuclei, we derive the equations of motion for the momentum amplitudes of each orbital. They contain an additional term compared with the single-electron case.
While we use the infinite-range exterior complex scaling (irECS) \cite{Scrinzi_2010,Orimo_2018} as an efficient absorbing boundary,
the present tSURFF implementation also supports other absorption boundaries such as mask functions and complex absorbing potentials.
We achieve highly accurate calculations of angle-resolved PES with considerably reduced computational costs.


This paper is organized as follows. 
We briefly review the TD-CASSCF and TD-ORMAS methods in Sec.~\ref{sec:td-ormas} and the tSURFF method for single-electron systems in Sec.~\ref{sec:tsurff}.
In Sec.~\ref{sec:tsurff_ormas}, we describe our theoretical application and numerical implementation of tSURFF to the TD-MCSCF method.
Numerical examples are presented in Sec.~\ref{sec:numerical_applicaiton}.
Conclusions are given in Sec.~\ref{sec:summary}.
We use Hartree atomic units unless otherwise indicated.

\section{the TD-CASSCF and TD-ORMAS methods\label{sec:td-ormas}}
We consider an $N$-electron atom with $N_{\uparrow}$ up-spin and $N_{\downarrow}$ down-spin electrons ($N = N_{\uparrow} + N_{\downarrow}$) in a laser field ${\bf E}(t)$. 
Its dynamics is described within the dipole approximation by the Hamiltonian,
\begin{equation}
    H(t) = H_1(t) + H_2,
\end{equation}
with,
\begin{align}
    H_1(t) &= \sum_{i=1}^N h(\boldsymbol{r}_i, \nabla_i, t) \notag\\ 
    &= \sum_{i=1}^N \left( -\frac{ \nabla_i^2}{2} - \frac{Z}{|\boldsymbol{r}_i|} - i\bold{A}(t) \cdot \nabla_i \right),\\
    H_2 &= \sum_{i=1}^N \sum_{j > i}^N \frac{1}{|\boldsymbol{r}_i - \boldsymbol{r}_j|},
\end{align} 
where $Z$ denotes the atomic number and $\boldsymbol{A}(t) = -\int_{-\infty}^t \boldsymbol{E}(t^\prime) dt^\prime$ the vector potential.
We use the velocity gauge, since ECS works only with it, and not with the length gauge \cite{McCurdy_1991}.

In the TD-CASSCF and TD-ORMAS methods, the total wave function is given by,
\begin{equation}
	\Psi (t) = \hat{A}\left[\Phi_{\rm fc}\Phi_{\rm dc}(t)\sum_I\Phi_I(t) C_I(t)\right],
\end{equation}
where $\hat{A}$ denotes the antisymmetrization operator, $\Phi_{\rm fc}$ and $\Phi_{\rm dc}$ the closed-shell determinants formed with $n_{\rm fc}$ frozen-core and $n_{\rm dc}$ dynamical-core orbitals, respectively, and $\{\Phi_I\}$ the determinants constructed from  $n_{\rm a}$ active orbitals. 
Whereas the active electrons are fully correlated among the active orbitals in the TD-CASSCF method, the TD-ORMAS method \cite{Sato_2015} further subdivides the active orbitals into an arbitrary number of subgroups, specifying the minimum and maximum number of electrons accommodated in each subgroup.  

The equations of motion (EOMs) describing the temporal evolution of the CI coefficients $\{C_I\}$ and the orbitals $\{\psi_p\}$ are derived on the basis of the time-dependent variational principle (TDVP) \cite{Moccia_1973,Sato_2015} and read,
\begin{gather}
    \label{eomci}
    i \frac{d}{dt} C_I(t) = \sum_J \braket{\Phi_I| \hat{H} - \hat{R}| \Phi_J}\\
    \label{eomorb}
	i \frac{d}{dt} \ket{\psi_p} = \hat{h} \ket{\psi_p} + \hat{Q} \hat{F}  \ket{\psi_p} + \sum_{q} \ket{\psi_q}  R^q_p,
\end{gather}
where $\hat{Q} = 1 - \sum_{q} \Ket{\psi_q} \Bra{\psi_q}$ the projector onto the orthogonal complement of the occupied orbital space.
$\hat{F}$ is a non-local operator describing the contribution from the interelectronic Coulomb interaction, defined as 
\begin{gather}
	\hat{F} \ket{\psi_p} = \sum_{oqsr} (D^{-1})^o_p P^{qs}_{or} \hat{W}^r_s \ket{\psi_q},
\end{gather}
where $D$ and $P$ are the one- and two-electron reduced density matrices, and $\hat{W}^r_s$ is given, in the coordinate space, by
\begin{gather}
	W^{r}_{s} \left(\boldsymbol{r} \right) = \int d \boldsymbol{r}^\prime \frac{\psi_{r}^{*} (\boldsymbol{r}^\prime) \psi_{s} ( \boldsymbol{r}^\prime )}{| \boldsymbol{r} - \boldsymbol{r}^\prime | } .
	\label{eq:W}
\end{gather}
The matrix element $R^q_p$ is given by,
\begin{gather}
\label{eq:orbital-time-derivative}
	R^q_p = i \braket{\psi_q | \dot{\psi_p} } - h^q_p,
\end{gather}
with $h^q_p = \braket{\psi_q|\hat{h}|\psi_p}$. 
$R^q_p$'s within one orbital subspace (frozen core, dynamical core and each subdivided active space) can be arbitrary Hermitian matrix elements, and in this paper, they are set to zero. 
On the other hand, the elements between different orbital subspaces are determined by the TDVP. Their concrete expressions are given in ref.~\cite{Sato_2015}, where $iX^q_p = R^p_q + h^q_p$ is used for working variables.

\section{the $\textrm{t}$SURFF method for single-electron systems\label{sec:tsurff}}
In this section, we briefly review the tSURFF method \cite{Tao_2012} for single-electron systems governed by TDSE,
\begin{gather} 
	i\frac{\partial}{\partial t} \Psi(\boldsymbol{r}, t) = h_1(t) \Psi(\boldsymbol{r}, t),\\
	h_1(t) = -\frac{1}{2} \nabla^2 + V(\boldsymbol{r}) - i \boldsymbol{A}(t) \cdot \nabla,
\end{gather}
where $V(\boldsymbol{r})$ denotes the nuclear potential.
The photoelectron momentum amplitude (PMA) $a(\boldsymbol{k}, t)$ for momentum $\boldsymbol{k}$ and PES
\begin{gather}
	\label{pes}
	\rho(E, t) = \int d\Omega |a(\boldsymbol{k}, t)|^2 |\boldsymbol{k}|^2 \quad (E = {|\boldsymbol{k}|^2}/2),
\end{gather}
can in principle be approximately calculated by projecting the outgoing wave packet residing outside a given radius $R_s$ onto the plane wave $(2\pi)^{-3/2}\exp (i\boldsymbol{k}\cdot\boldsymbol{r})$ at a time $t$ sufficiently after the pulse.
Then, PMA is given by,
\begin{align}
	\label{pma}
	a(\boldsymbol{k}, t) &= \braket{\chi_{\boldsymbol{k}}(\boldsymbol{r}, t) |\hat{\theta}(R_{\rm s})| \Psi(t)}, \\
	&\equiv \int \chi_{\boldsymbol{k}}^*(\boldsymbol{r},t)\theta(|\boldsymbol{r}|-R_s)\Psi (\boldsymbol{r},t)d^3\boldsymbol{r}, \\
	&= \int_{r>R_s}\chi_{\boldsymbol{k}}^*(\boldsymbol{r},t)\Psi(\boldsymbol{r},t)d^3\boldsymbol{r}.
\end{align}
where $\chi_{\boldsymbol{k}}(\boldsymbol{r},t)$ denotes the Volkov wave function and $\theta (x)$ the Heaviside function.


On the other hand, the tSURFF method calculates PES by a time integration of the wave function surface flux, based on an assumption that the nuclear potential does not affect the time evolution of the distant photoelectron wave packet. 
Under this assumption, the Volkov wave functions and photoelectron wave packets in the region $|\boldsymbol{r}| > R_{\rm s}$ are evolved by the same nuclear-potential-free Hamiltonian
\begin{gather}
	h_{\rm s} = -\frac{1}{2}\nabla^2 - i \boldsymbol{A}(t) \cdot \nabla.
\end{gather}
By taking time derivative of Eq.~\eqref{pma}, we obtain the EOM of the momentum amplitude,
\begin{align}
	\label{eom1e_tsurff}
	-i \frac{\partial}{\partial t} a(\boldsymbol{k}, t) 
	&=  \braket{\chi_{\boldsymbol{k}}(t) | [h_{\rm s}, \theta(R_{\rm s})] | \Psi(t)}.
\end{align}
As shown in Ref.~\cite{Tao_2012}, since all the terms appearing in the commutator $[h_{\rm s}, \theta(R_{\rm s})]$ contain delta functions $\delta(r-R_{\rm s})$ [see Eq.~\eqref{commutator} below], wave functions and their spatial derivative only on the surface $|\boldsymbol{r}| = R_{\rm s}$ are required to solve Eq.~\eqref{eom1e_tsurff}. Hence, it is no longer needed to keep the whole wave function and allowed to use an absorbing boundary, which leads to a significant computational cost reduction.


\section{Application of the tSURFF method to the TD-ORMAS simulations \label{sec:tsurff_ormas}}
\subsection{Photoelectron reduced density matrix}
To obtain PES in multielectron systems described by the multiconfiguration expansion Eq.~(\ref{eq:multiconfiguration expansion}), we define the photoelectron reduced density matrix (PRDM).
Since our definition of ionization is based on the spatial domain $|{\bf r}|>R_s$ , the single particle reduced density matrix of a photoelectron in the coordinate space can be defined as,
\begin{align}
	P(\boldsymbol{r}, \boldsymbol{r'}) = \sum_{pq} \bra{\boldsymbol{r}}\theta(R_s)\ket{\psi_p} 
	D^p_q \bra{\psi_q}\theta(R_s)\ket{\boldsymbol{r'}}.
\end{align}
In the momentum space, its elements are given by,
\begin{multline}
	\label{momentum_PRDM}
	\tilde{P}(\boldsymbol{k}, \boldsymbol{k'}) = \int d\boldsymbol{r}d\boldsymbol{r'} \braket{\chi_{\boldsymbol{k}}(t)|\boldsymbol{r'}} P(\boldsymbol{r}, \boldsymbol{r'}) \braket{\boldsymbol{r}|\chi_{\boldsymbol{k}}(t)} \\
	=  \sum_{pq} \bra{\chi_{\boldsymbol{k}}(t)} \theta(R_s)\ket{\psi_p}
	D^p_q \bra{\psi_q}\theta(R_s)\ket{\chi_{\boldsymbol{k'}}(t)}.
\end{multline}
The diagonal part $\tilde{P}(\boldsymbol{k}, \boldsymbol{k})$ is interpreted as photoelectron momentum distribution, and, then, PES is given by,
\begin{equation}
	\rho(E) = \int d\Omega \tilde{P}(\boldsymbol{k}, \boldsymbol{k}) |\boldsymbol{k}|^2.
\end{equation}
A similar definition of the PRDM as Eq.~(\ref{momentum_PRDM}) and the direct projection onto scattering states were used 
to compute photoelectron spectrum in Ref.~\cite{Omiste_2017}.


\subsection{EOMs of momentum amplitudes of orbitals }

In this subsection, we derive the EOM of the momentum amplitude of orbital $p$,
\begin{gather}
	a_p(\boldsymbol{k}, t) = \braket{\chi_{\boldsymbol{k}}(t) |\theta(R_{\rm s})| \psi_p},
\end{gather}
which appears in Eq.~\eqref{momentum_PRDM}.
We assume that the nuclear potentials are negligible for photoelectrons as in the single-electron case and additionally that the interelectronic Coulomb interaction does not affect the dynamics of photoelectrons in the region beyond the radius $R_{\rm s}$.
Then, the orbital EOM Eq.~\eqref{eomorb} can be approximated as, 
\begin{gather}
	\frac{d}{dt} \ket{\psi_p} \simeq -i \left[ \hat{h}_{\rm s} \ket{\psi_p} - \sum_q \ket{\psi_q} \left\{\bra{\psi_q} \hat{F}  \ket{\psi_p} -  R^q_p \right\} \right],
\end{gather}
so that EOMs of $a_p(\boldsymbol{k})$ is obtained as
\begin{align}
	\label{eomme_tsurff}
	-i \frac{\partial}{\partial t} a_p(\boldsymbol{k}, t) =& \braket{\chi_{\boldsymbol{k}}(t) | [h_{\rm s}, \theta(R_{\rm s})] | \psi_p(t)} \notag\\
	+& \sum_q a_q(\boldsymbol{k}, t) \{\bra{\psi_q(t)} \hat{F} \ket{\psi_p(t)} - R^q_p\}.
\end{align}
It should be noted that the second term in Eq.~\eqref{eomme_tsurff} represents a significant difference from the single-electron case [Eq.~(\ref{eom1e_tsurff})]. This term includes the effect of the interelectronic Coulomb interaction inside $R_s$ and is not negligible; for example, the phase variation due to the energy of the ionic core is reflected in photoelectron momentum spectra through this term.


It may not be a priori obvious if the Coulomb interaction between electrons in the outer region is negligible. 
A numerical validation of this approximation will be given in Sec. \ref{sec:numerical_applicaiton}.

\subsection{Implementation}
In this subsection, we describe the implementation of the tSURFF method to TD-MCSCF simulations. The implementation is based on our TD-CASSCF code \cite{Sato_2016} for atoms subject to a laser pulse linearly polarized in the $z$ direction. 
The orbitals are discretized with spherical finite-element discrete-variable-representation (FEDVR) basis functions \cite{McCurdy_2004},
\begin{gather}	
	\label{discretized_orb}
	\psi_p(r, \theta, \phi) = \sum_{kl} c^p_{kl} f_k(r) Y_{lm_p}(\theta, \phi),
\end{gather}
where $f_k(r)$ and $Y_{lm_p}(\theta, \phi)$ denote the FEDVR function and the spherical harmonics, respectively. 
The magnetic quantum number $m_p$ for orbital $\psi_p$ does not change due to the $z$-polarization of the laser pulse \cite{Sato_2016}.
As an absorbing boundary, infinite-range exterior complex scaling is implemented \cite{Scrinzi_2010, Orimo_2018}, which shows high efficiency and accuracy.
On the other hand, we discretize the orbital momentum amplitudes with grid points in the spherical coordinates,
where the Volkov wave function for a momentum $\boldsymbol{k} = (k, \theta_k, \psi_k)$ is given by,
\begin{align}
	\label{volkov}
	\chi_{\boldsymbol{k}}(\boldsymbol{r}, t) &= \frac{ \exp(-i\Lambda(\boldsymbol{k}, t))}{(2\pi)^{3/2}} \notag \\
	&\times \sum_{lm} 4\pi i^l Y^*_{lm}(\theta_k, \psi_k)  j_l(kr)Y_{lm}(\theta, \psi),
\end{align}
with  $j_l(kr)$ being the spherical Bessel function of the first kind and $\Lambda(\boldsymbol{k},t)$ the Volkov phase given by,
\begin{gather} 
	\label{volkov_phase}
	\Lambda(\boldsymbol{k},t) = \int_0^t \frac{1}{2} [\boldsymbol{k} - \boldsymbol{A}(\tau)]^2 d\tau.
\end{gather}

The commutator in Eq.~\eqref{eom1e_tsurff} can be rewritten as,
\begin{align}
	\label{commutator}
	 [h_{\rm s}, \theta(R_{\rm s})] &= -\frac{1}{2} \left[ \frac{\partial}{\partial r}\delta(r-R_{\rm s}) + \delta(r-R_{\rm s}) \frac{\partial}{\partial r} \right] \notag\\ 
	&- \frac{\delta(r-R_{\rm s})}{r} + i A_z(t) \cos(\theta) \delta(r-R_{\rm s}),
\end{align}
with $z$ component $A_z(t)$ of the vector potential $\boldsymbol{A}(t)$.
Using Eqs.~\eqref{volkov} and \eqref{volkov_phase} and introducing $g^p_{l}(r) = \sum_k  c^p_{kl} f_k(r)$, we can decompose the first term of Eq.~\eqref{eomme_tsurff} into,
\begin{widetext}
\begin{multline}
	\braket{\chi_{\boldsymbol{k}}(t) | [h_{\rm s}, \theta(R_{\rm s})] | \psi_p(t)} = 
	\frac{\exp{i\Lambda(\boldsymbol{k}, t)}}{(2\pi)^{3/2}} 4\pi \sum_l (-i)^l \Big[ Y_{lm_p}(\theta_k, \psi_k) \frac{ R_{\rm s}^2}{2} \{ j^*_l{}'(kR_{\rm s})g^p_l(R_{\rm s})  
	- j^*_l(kR_{\rm s})g^p_l{}'(R_{\rm s}) \} \\
	+ iA_z (t) \sum_{l'}(-i)^{l'} Y_{l'm_p}(\theta_k, \psi_k)  ) R_{\rm s}^2  j^*_l(kR_{\rm s})g^p_l(R_{\rm s})  \alpha_{l'lm} \Big],
\end{multline}
\end{widetext}
where $ j^*_l{}'(r)$ and $g^p_l{}'(r)$ denote the radial derivative of $ j^*_l(r)$ and $g^p_l(r)$, respectively.
For the second term in Eq.~\eqref{eomme_tsurff}, since $\{\bra{\psi_q(t)} \hat{F} \ket{\psi_p(t)} - R^q_p\}$ is evaluated during the orbital propagation [Eq.~(\ref{eomorb})], we can reuse it.

We integrate Eq.~\eqref{eomme_tsurff} with the first order exponential integrator \cite{Cox_2002} by treating Eq.~\eqref{eomme_tsurff} as simultaneous inhomogeneous linear differential equations. The evolution of a vector $\boldsymbol{a}(\boldsymbol{k}, t) \equiv \{a_p(\boldsymbol{k}, t) \}$  from the time $t$ to $t+\Delta t$ is described as 
\begin{multline}
    \boldsymbol{a}(\boldsymbol{k}, t+\Delta t) = \boldsymbol{a}(\boldsymbol{k}, t) \exp[i \Delta t X(t)]   \\
	+ \boldsymbol{S}(t) \frac{\exp[i \Delta t X(t)]  - 1}{X(t)} ,
\end{multline}
where the matrix $X(t) = \{X_{qp}(t)\} $ and vector $\boldsymbol{S}(t) = \{S_p(t)\}$ are defined as, 
\begin{alignat}{1}
	X_{qp}(t) &= \bra{\psi_q(t)} \hat{F} \ket{\psi_p(t)} - R^q_p, \\
	S_p(t) &= \braket{\chi_{\boldsymbol{k}}(t) | [h_{\rm s}, \theta(R_{\rm s})] | \psi_p(t)}.
\end{alignat}
We compute $\exp[i \Delta t X(t)]$ by directly diagonalizing $X(t)$ at every time step, which is not demanding since the size of $X(t)$ is $[N_{\rm orb} \times N_{\rm orb}]$ and the number of orbitals $N_{\rm orb}$ usually falls within the range from a several to several tens.
\section{Numerical results\label{sec:numerical_applicaiton}}
In this section, we present numerical applications of the implementation of tSURFF to the TD-MCSCF method described in the previous section.
The electric field of the laser pulse is assumed to have the following shape for simulations of Ne atom (Sec. \ref{sec:Ne}):
\begin{gather}
    E(t) = \sqrt{I_0} \sin^2 \left( \pi \frac{t}{N_{\rm opt}T} \right) \sin \omega t ,\   0 \leq t \leq N_{\rm opt}T,
	\label{eq:laser}
\end{gather}
and for Ar atom (Sec. \ref{sec:Ar}):
\begin{gather}
    E(t) = e(t) \sin\omega t, \\
    e(t) = 
    \begin{cases}
    \displaystyle \sqrt{I_0} \frac{t}{2T} ,&  0 < t \leq 2T\\[8pt]
    \displaystyle \sqrt{I_0} ,& 2T < t \leq (N_{\rm opt} -2) T\\[8pt]
    \displaystyle \sqrt{I_0} \frac{N_{\rm opt}T - t}{2T} ,&  (N_{\rm opt} -2) T < t \leq N_{\rm opt} T
    \end{cases},
	\label{eq:laser}
\end{gather}
where $I_0$ is a peak intensity, $T$ is a period at the central frequency $\omega = 2\pi/T$, and $N_{\rm opt}$ is the total number of optical cycles.

\subsection{Ne atom \label{sec:Ne}}
We first calculate the PES of a neon atom subject to an attosecond pulse with a peak intensity of $2.5 \times 10^{12} \rm{W/m}^2$, a wavelength of 12.398 nm corresponding to 100 eV photon energy, and $N_{\rm opt}=16$.
The results by tSURFF and direct projection on plane waves are compared.
As an absorbing boundary we use irECS with a scaling radius $R_0$ of 40 a.u for tSURFF and 400 a.u for direct projection. The latter is large enough to hold the departing wave packet from two-photon ionization. 
$R_s=40\,{\rm a.u.}$ for tSURFF, and the wave packet outside this radius is used for projection.

\begin{figure}[tb]
      \begin{minipage}[t]{1\hsize}
      	\includegraphics[width=1.0\linewidth, clip, bb = 0 0 288 144]{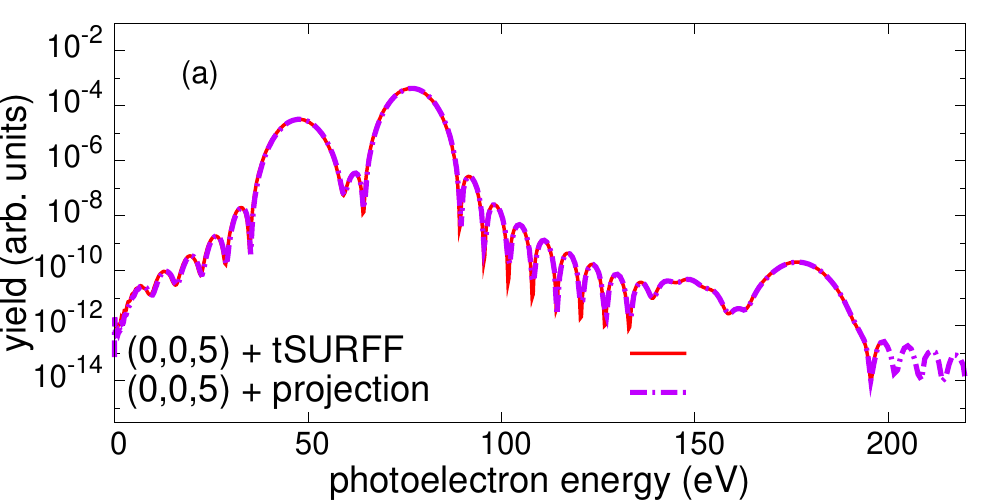}
      	\label{Ne_005}
      \end{minipage} \\
      \begin{minipage}[t]{1\hsize}
      	\includegraphics[width=1.0\linewidth, clip, bb = 0 0 288 144]{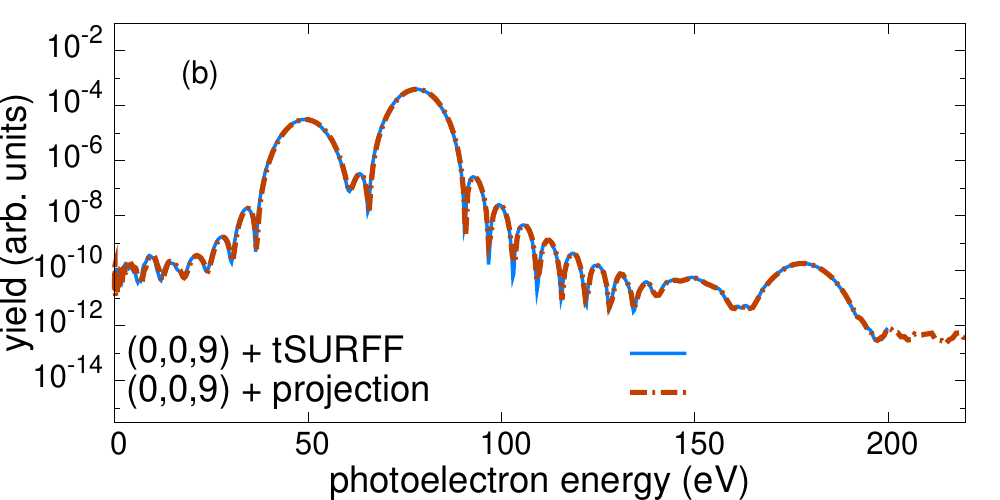}
	    \label{Ne_009}
      \end{minipage}
      \begin{minipage}[t]{1\hsize}
      	\includegraphics[width=1.0\linewidth, clip, bb = 0 0 288 144]{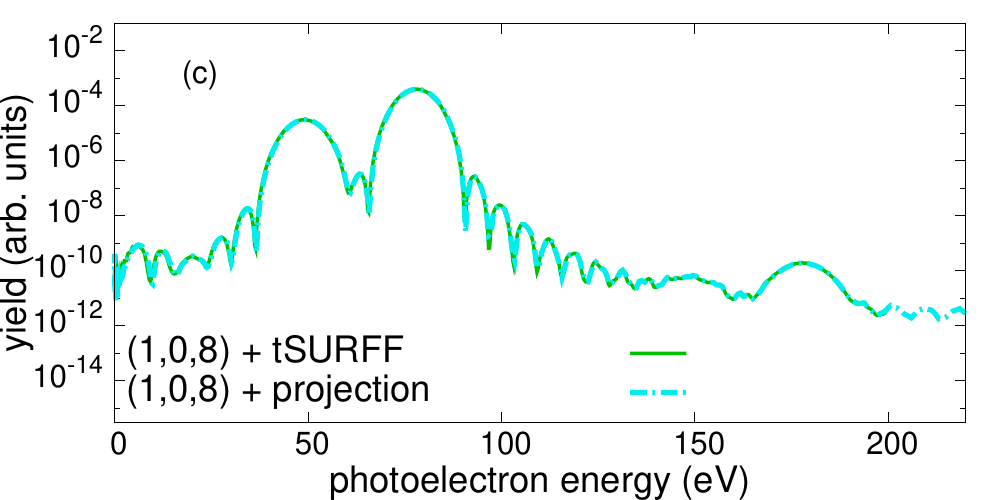}
	    \label{Ne_108}
      \end{minipage}
      \caption{Photoelectron energy spectra of a Ne atom subject to an attosecond pulse with 100eV photon energy calculated with orbital classification (a) $(n_\text{fc}, n_\text{dc}, n_\text{a}) = (0, 0, 5)$ (TDHF), (b) $(0, 0, 9)$ (MCTDHF), and (c) $(1, 0, 8)$ (TD-CASSCF). The results by tSURFF and direct projection are compared.}
      \label{Ne_100eV}
\end{figure}

We do TD-CASSCF simulations with 3 kinds of orbital classifications $(n_\text{fc}, n_\text{dc}, n_\text{a}) = (0, 0, 5)$, $(0, 0, 9)$, and $(1, 0, 8)$, where $n_\text{fc}$, $n_\text{dc}$, and $n_\text{a}$ are the number of frozen-core, dynamical-core, and active orbitals, respectively. Note that the first and second correspond to the time-dependent Hartree-Fock (TDHF) \cite{Dirac_1930} and MCTDHF methods, respectively.
The results are shown in Fig.~\ref{Ne_100eV}.
We see single photon ionization peaks (around 30 - 90 eV) and two photon above threshold ionization (ATI) peaks (around 130 eV - 190eV) from $2s$ and $2p$ orbitals.
The agreement between the results by tSURFF and direct projection is excellent. This shows the validity of the neglect of the electron-electron and nucleus-electron Coulomb interaction beyond $R_s$ assumed in tSURFF.
While the single photon ionization peaks from $2s$ and $2p$ orbitals are expected to be at 51.5 and 78.4 eV, respectively, based on the experimental values of the binding energies \cite{NIST}, the peaks in the calculated spectra are located at 47.5 and 76.7 eV in Fig.~\ref{Ne_100eV}(a), 48.9 and 77.9 eV in (b), and 48.7 and 77.9 eV in (c), coming closer to the experimental positions with increasing number of orbitals. 
Otherwise, TDHF gives results similar to the MCTDHF and TD-CASSCF ones for this process.

\subsection{Ar atom \label{sec:Ar}}

\begin{figure}[tb]
    \begin{minipage}[h]{1.0\hsize}
        \begin{flushright}
    	\includegraphics[width=1.0\linewidth, clip, bb = 0 0 288 180]{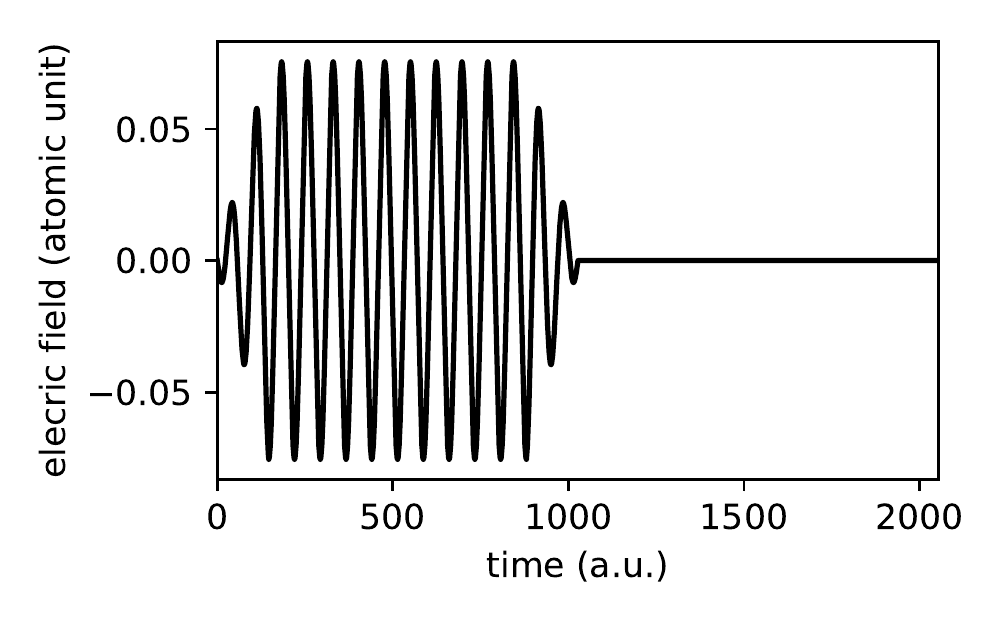}
    	\end{flushright}
	\end{minipage}\\
    \begin{minipage}[h]{1.0\hsize}
        \begin{flushright}
    	\includegraphics[width=0.95\linewidth, clip, bb = 0 0 288 180]{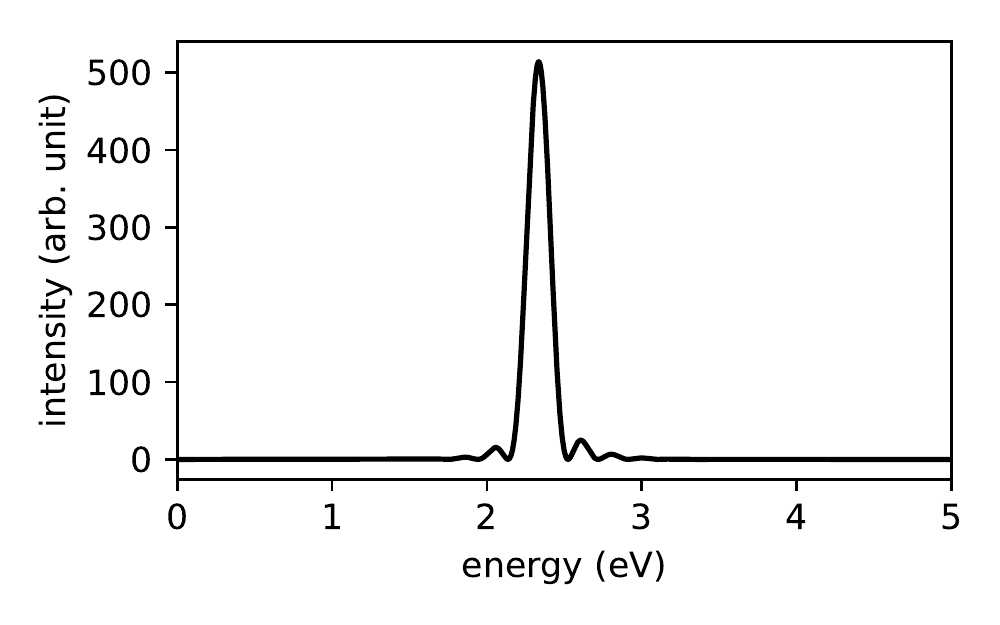}
    	\end{flushright}
	\end{minipage} \
	\caption{Temporal shape and intensity spectrum of the laser pulse considered for the case of Ar.\label{ar_laser}}
\end{figure}

Next, we simulate ATI of a Ar atom subject to an intense visible laser pulse using the TD-ORMAS method and discuss the effect of electronic correlation. 
We consider a pulse, which has a peak intensity of $2.0 \times 10^{14}\,\mathrm{W/cm^2}$, a wavelength of 532 nm, and a pulse width of $N_{\rm opt}=14$ optical cycles. The ponderomotive energy $U_p$ is 5.285 eV.
The temporal shape and energy spectrum of the pulse are shown in Fig.~\ref{ar_laser}.

\begin{figure}[tb]
	\includegraphics[width=0.85\linewidth, clip, bb = 0 0 274 293]{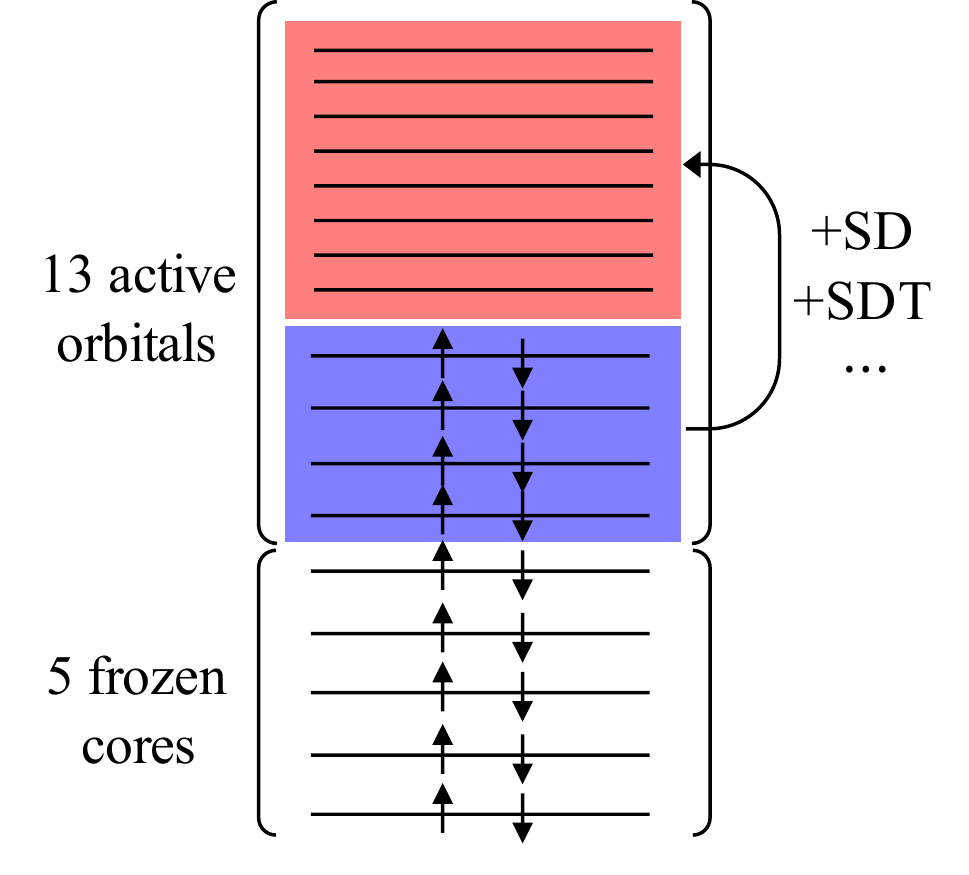}	
	\caption{An example of the orbital subdivision of the TD-ORMAS method for an Ar atom with 18 electrons, which wave function is composed of 5 frozen cores and 13 active orbitals, $(n_\text{c}, n_\text{d}, n_\text{a}) = (5,0,13)$. 
	The excitation restrictions SD (SDT, $\cdots$) indicates that single and double (single, double and triple, $\cdots$) excitation from the blue to red group are only allowed.}
	\label{ormas_example}
\end{figure}

Here, we subdivide $n_\text{a}$ active orbitals into 4 and $(n_\text{a} - 4)$ orbitals, as schematically illustrated in Fig.~\ref{ormas_example} for orbital decomposition $(n_\text{c}, n_\text{d}, n_\text{a}) = (5,0,13)$. 
By setting the maximum number of electrons in the second subgroup [$(n_\text{a} - 4)$ active orbitals] to 2 or 3, the configurations with up to double (SD) or triple (SDT) excitation are considered, respectively.

\begin{figure}[tb]
	\includegraphics[width=1.0\linewidth, clip, bb = 0 0 288 432]{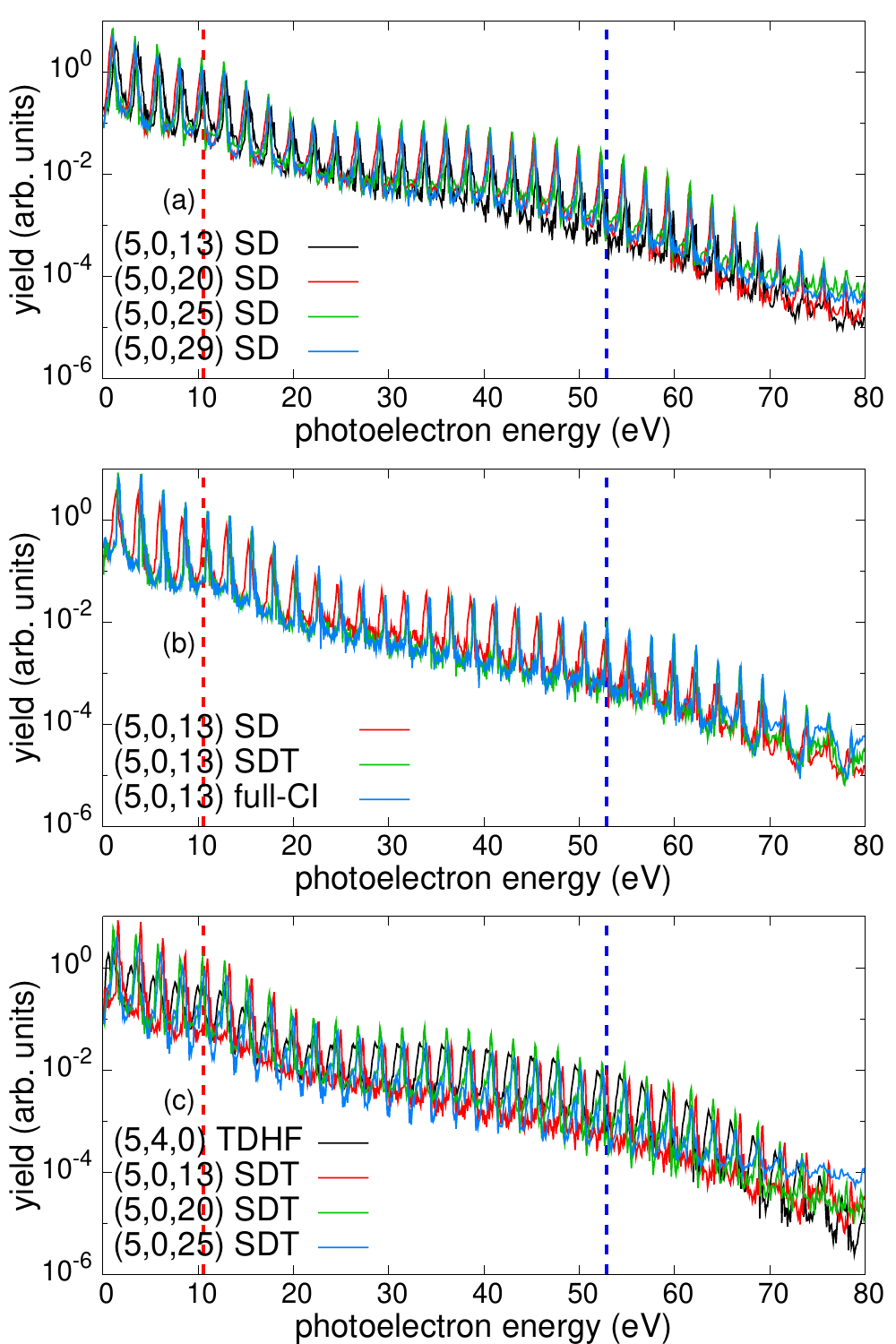}
	\caption{Photoelectron energy spectra of an Ar atom subject to a visible intense laser pulse with a wavelength of 532 nm and an intensity of $2.0 \times 10^{14} \mathrm{~W/cm^2}$. The red and blue dashed vertical lines show $2U_p$ and $10U_p$ ($U_p=5.285$ eV). The results with different number of orbitals $(n_\text{fc}, n_\text{dc}, n_\text{a})$ and excitation restrictions (SD, SDT, full-CI) are compared.}
	\label{ar_espec}
\end{figure}

\begin{figure}[tb]
    \includegraphics[width=1.0\linewidth, clip, bb = 0 0 356 556]{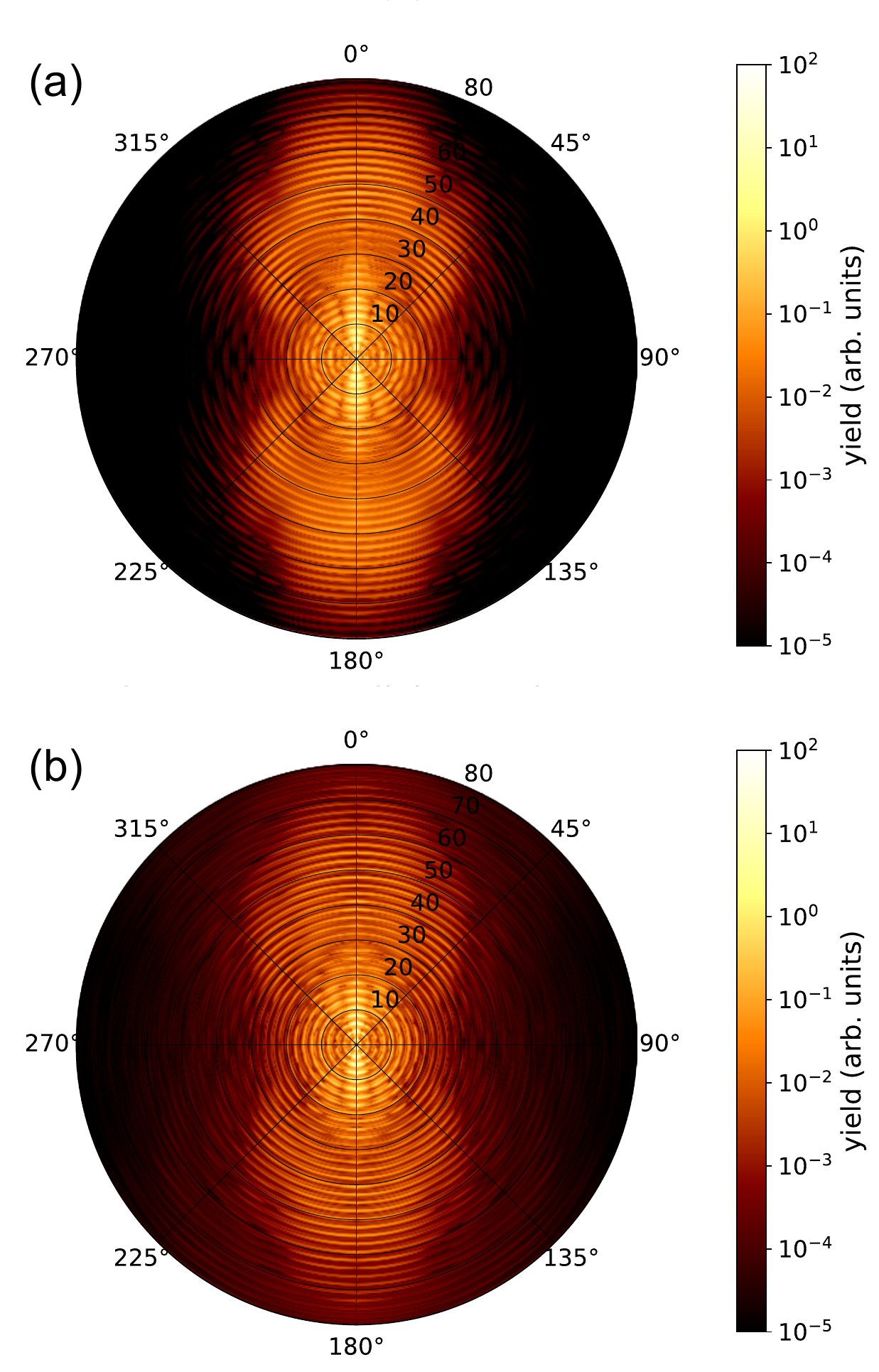}
	\caption{Angle-resolved photoelectron energy spectra of a Ar atom subject to a visible intense laser pulse with a wavelength of 532 nm and an intensity of $2.0 \times 10^{14} \mathrm{~W/cm^2}$. The laser polarization ($z$ direction) corresponds to $0^\circ$. (a) $(n_\text{fc}, n_\text{dc}, n_\text{a}) = (5, 4, 0)$, i.e., TDHF (b) $(n_\text{fc}, n_\text{dc}, n_\text{a}) = (5, 0, 25)$ and SDT excitation restriction.}
	\label{ar_adpes}	
\end{figure}

\begin{figure}[tb]
	\includegraphics[width=1.0\linewidth, clip, bb = 0 0 288 432]{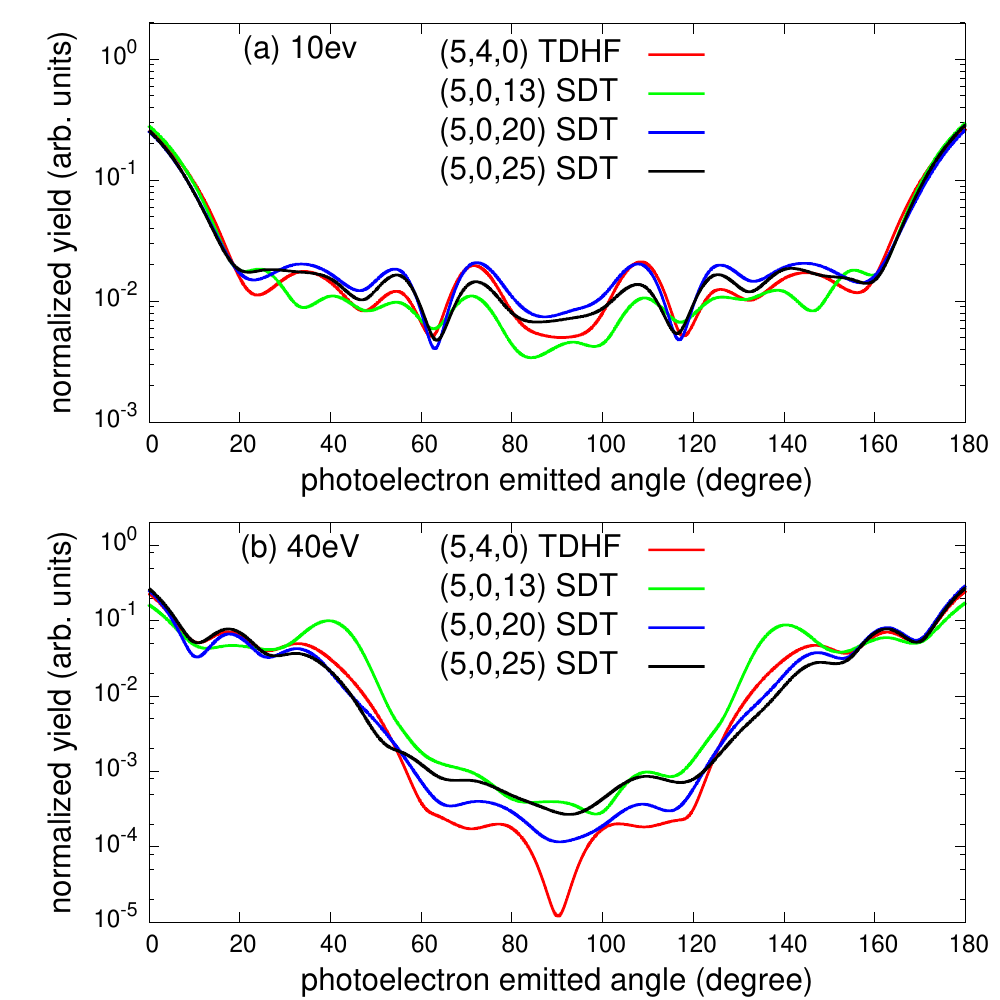}
	\caption{Angular distribution of photoelectron yields at (a) 10 eV and (b) 40 eV, averaged over a $\pm1.1 \text{eV}$ energy range. The angle-integrated yield is normalized to unity. The results with different orbital conditions are compared.}
	\label{arpes_ev}
\end{figure}

Figure~\ref{ar_espec} shows PES calculated with different orbital classifications. 
We can recognize the direct cutoff at $2U_p$ and rescattering cutoff at $10U_p$.
In Fig.~\ref{ar_espec}(a), the results with different numbers of active orbitals within single and double (SD) excitation are compared. 
The spectrum is nearly converged with 25 and 29 active orbitals.
Figure~\ref{ar_espec}(b) compares the results with SD and SDT excitation restriction and full CI (TD-CASSCF), with 13 active orbitals. The SDT and full CI results almost overlap each other, indicating that SDT is sufficient for numerical convergence.
Thus, the result using 25 active orbitals with SDT excitation is expected to be numerically nearly exact. Then, in Fig.~\ref{ar_espec}(c), we compare the PES calculated using 13, 20 and 25 active orbitals with SDT excitation restriction and also the TDHF result. The peak positions slightly depend on the number of orbitals, as we have also seen in Fig.~\ref{Ne_100eV}. Moreover, in the TDHF case, the peaks are significantly broadened. 
The ATI peak position $E_n$ corresponding to $n$-photon absorption is given by,
\begin{equation}
	E_n = n \hbar\omega - I_{\rm p} - U_{\rm p},
\end{equation}
where $I_{\rm p}$ is the ionization potential. The difference in peak position observed in Fig.~\ref{ar_espec} can be attributed to that in $I_{\rm p}$, which depends on the number of orbitals and excitation restriction.
In addition, in mean-field approaches such as TDHF, the ionization potential effectively increases as ionization proceeds and the electron density near the nucleus decreases \cite{Kulander_1992}. This results in the peak broadening.

Finally, in Fig.~\ref{ar_adpes}, we compare ARPES calculated with the TD-ORMAS method using 25 active orbitals with SDT excitation restriction and with the TDHF method using 4 dynamical-core orbitals.
We see difference in detailed structure. 
In particular, the high-order ($>2U_p$) rescattering contribution has much broader angular distribution in the TD-ORMAS result. 
The difference can also be clearly observed in Fig~\ref{arpes_ev}, which shows the photoelectron angular distribution (PAD) at 10 eV ($<2U_p$) and 40 eV ($>2U_p$), representatives of the lower and higher energy regions, respectively.
At 10 eV photoelectron energy, where the main contribution is from direct ionization, all the results exhibit similar behavior. 
In contrast, at 40 eV photoelectron energy, for which rescattering from the parent ion is involved and thus strong electron correlation is expected, the calculated PAD varies with the number of orbitals till it approximately converges with $n_{\rm a}=25$.
Especially, the TDHF method significantly underestimates the yield in the direction ($90^\circ$) perpendicular to the laser polarization.
This indicates that electronic correlation is non-negligible in detailed discussion of ATI ARPES.

%




\section{Summary\label{sec:summary}}

We have presented a successful numerical implementation of tSURFF to the TD-MCSCF (TD-CASSCF and TD-ORMAS) methods to extract angle-resolved photoelectron energy spectra from laser-driven multielectron atoms.
We have derived the EOMs for photoelectron momentum amplitudes of each orbital.
To obtain PES in systems described within the MCSCF framework, the photoelectron reduced density matrix has been introduced, whose diagonal elements in the momentum space correspond to PES.
Since one of the biggest benefits of tSURFF is no need to hold the complete wave function within the simulation box, it allows combined use of an efficient absorbing boundary such as irECS \cite{Scrinzi_2010,Orimo_2018}.


We have applied the present implementation to Ne and Ar atoms subject to attosecond XUV pulses and intense visible laser pulses, respectively.
We have demonstrated converged calculation of ATI spectra from Ar including electronic correlation, which would be prohibitive without tSURFF and irECS. 
It is expected that this achievement enables direct comparison with experiments and precise prediction of high-field and ultrafast phenomena.

While we have presented the application of tSURFF to TD-MCSCF methods in this study, it is straightforward to extend it to other multielectron {\it ab initio} methods using time-dependent orbitals such as the time-dependent optimized coupled-cluster method \cite{Sato_2018} and TD-MCSCF methods including nuclear dynamics \cite{Anzaki_2017}.
Such applications would enable us to compute ARPES from even more complicated systems and processes.

\begin{acknowledgments}
This research was supported in part by a Grant-in-Aid for Scientific Research 
(Grants No.~16H03881, No.~17K05070, and No.~18H03891)
from the Ministry of Education, Culture, Sports, Science and Technology (MEXT) of Japan and also 
by the Photon Frontier Network Program of MEXT.
This research was also partially supported by the Center of Innovation Program from the Japan Science 
and Technology Agency (JST), by CREST (Grant No.~JPMJCR15N1), JST, by Quantum Leap Flagship Program of MEXT, and by JSPS and HAS under the Japan-Hungary Research Cooperative Program.
Y.~O. gratefully acknowledges support from the Graduate School of
Engineering, The University of Tokyo, Doctoral Student Special
Incentives Program (SEUT Fellowship).
\end{acknowledgments}

\bibliography{all10}

\end{document}